\documentstyle[12pt]{article}
\def\graphic #1#2#3#4#5{

\noindent
\hrulefill
    \vskip#2 \relax
    \vskip -3.9 cm
    \hskip 4.8 cm
    {\large \bf Universidade do Estado do Rio de Janeiro }
    \newline

    \vskip -0.25 cm
    \hskip 7.5 cm
    {\large \bf Instituto de F{\'\i}sica }

 	\vskip 1 cm
    \hskip 7.5 cm
    {\large Phys-Pub #4 }

    \hskip 7.5 cm
    {\large Preprint}

    \hskip 7.5 cm
    {\large #5 }

\medskip
\noindent
\hrulefill

    \vskip 2.9 cm
    }
\def\ugraphic #1#2#3{
    \vskip#2 \relax
    \centerline{\hbox to#1{\special{bmp:#3 x=#1, y=#2}\hfil}}}
\begin{document}
\hspace\parindent
\thispagestyle{empty}

\graphic{2 in}{1.6 in}{uerj.bmp}{020/95}{September 1995}
\centerline{\LARGE \bf Analytical Solving of}
\medskip
\centerline{\LARGE \bf Partial Differential Equations}
\medskip
\centerline{\LARGE \bf using Symbolic Computing}


\bigskip
\bigskip
\bigskip
\centerline{\large
E.S.Cheb-Terrab\footnote{Departamento de F\'{\i}sica Te\'orica, IF-UERJ.
E-mail: terrab@vmesa.uerj.br}
K. von B\"ulow\footnote{Instituto de Matem\'atica Pura e Aplicada, IMPA.
E-mail: kathi@impa.br}
}

\bigskip
\bigskip
\bigskip
\bigskip
\begin{abstract}
This work presents a brief discussion and a plan towards the analytical
solving of Partial Differential Equations (PDEs) using symbolic computing,
as well as a implementation of part of this plan as the {\it PDEtools}
software-package of commands.
\end{abstract}

\bigskip
\bigskip
\centerline{ \underline{\hspace{6.5 cm}} }

\medskip
\centerline{ {\bf To be published in the Proceedings of }}

\centerline{ {\bf {\it Computing in High Energy Physics, CHEP-1995}}}

\centerline{{\bf World Scientific Publishing Co.}}

\newpage
\section*{Introduction}
The {\it PDEtools} software-package here presented is a MapleV R.3
implementation of analytical methods for solving and working with PDEs.
The main commands of the package are: {\bf pdsolve}, a  PDE-solver;  {\bf
dchange}, for making changes of variables; {\bf sdsolve}, for solving
systems of Ordinary Differential Equations (ODEs); {\bf splitsys}, for
splitting up systems of ODEs; and {\bf mapde}, for mapping PDEs into more
convenient PDEs. New features, with respect to the first version of the
package\cite{p10}, are: the {\bf sdsolve}, {\bf splitsys} and {\bf mapde}
commands, as well as significant improvements made to {\bf dchange}.

The solving methods implemented in this version are, mainly, standard
methods such as the characteristic strip for first-order PDEs, reduction
to canonical form of second-order PDEs, separation of variables,
etc.\cite{courant,smirnov}. Furthermore, the user can optionally
participate in the solving process by giving the solver an extra argument
(the \verb-HINT- option) indicating a {\it functional form} for the
indeterminate function. Note that, besides the inherent restrictions of
the methods used in this release, the {\it PDEtools} package cannot tackle
systems of PDEs.
\section{The strategy for a PDE-solver}
\subsection{The objectives of a PDE solving command}
\label{objectives}
The first point to be discussed is what the program ({\bf pdsolve}) should
return as the result. Our idea was that, given a PDE, {\bf pdsolve} should
return the most {\it general solution} it could find. At least, it should
try to separate all the variables, giving the user the option of asking
for the integration of the resulting {\it uncoupled} ODEs in order to
arrive at a {\it complete solution}. {\it General solutions} depend on $n$
arbitrary functions of $k-1$ variables, where $n$ is the differential
order of the given PDE and $k$ is the number of independent variables.
{\it Complete solutions}, on the other hand, depend on (at least)
$\sum_{i=1}^{k} n_i$ arbitrary constants, where $n_i$ represents the
differential order w.r.t each of the {\it k} variables\footnote{Depending
on the PDE, {\bf pdsolve} may also return a solution in some sense
``intermediate", containing less than $n$ arbitrary functions but more
general than a {\it complete} one.}.
\subsection{The solving methods and the plan}
\label{plan}
Let us classify the methods which may lead to the desired types of
solutions into {\it symmetry} methods - those relying on the use of the
theory of invariance under Lie groups of transformations\cite{olver} -, and
{\it standard} methods - the other ones\cite{courant,smirnov}.
Though it seems clear to us that a PDE-solver should make use of the
systematic {\it symmetry} methods, these methods usually map the task of
solving a PDE into a task of the same type, requiring the solving of the
system of PDEs that determines the generators of the invariance group. On
the other hand, {\it standard} methods permit the mapping of the task of
solving a PDE into the tasks of realizing algebraic manipulations,
evaluating integrals and solving ODEs, all of which are already
implemented in Maple. For these reasons,  we split the general objective
of solving PDEs analytically into: 1.\ the implementation of {\it
standard} methods for solving a single PDE; 2.\ the implementation of {\it
standard} methods for solving systems of coupled linear PDEs; 3.\ the
implementation of {\it symmetry} methods. This version of the {\it
PDEtools} package is concerned with the first step and does not make use
of {\it symmetry} methods.
\section{The PDE-solver: an approach using {\it standard} methods}
\label{standard}
{\it Standard} methods can be divided into those having a clear
prescription of when and how to be applied, e.g. the characteristic strip
method, and those which work in a more {\it heuristic} manner, such as
separation of variables. Also, all methods map the original problem into
another problem, and the latter still needs to be solved. Furthermore,
there are PDEs which can be tackled by more than one method, as well as
PDEs (the majority) for which we do not know any specific method of
solution.

Taking all these facts into account, it became clear to us that a
wide-range PDE-solver should be able to make decisions concerning the
method to be employed in {\it any} case, the trial strategy, and what to
do when the strategy followed at first fails (perhaps try another
method...). Therefore, our idea was to develop routines implementing {\it
deterministic} methods, to be used right at the beginning of the solving
process; and invest hard on algorithms for separating variables, to be
used when the implemented {\it deterministic} methods either are not
applicable or map the original PDE into a problem that the system fails to
solve.
\subsection{The {\it deterministic} methods}
\label{deterministic}
As mentioned above, we denominate {\it deterministic} a method having a
clear prescription of when and how to be applied. The methods falling into
this category which are implemented in this release of the {\it PDEtools}
package are just for first and second order PDEs. The most relevant ones
are: the characteristic strip method\footnote{A refinement was developed
in order to transform the parameterized solution of the strip, when
possible, into a explicit {\it general} solution using differential
invariants.}, the mapping between PDEs (realized using {\bf mapde}), and
some other particular methods applicable when the given PDE matches
a pattern recognized by {\bf pdsolve}. A detailed list of
the implemented methods can be seen in \cite{p10}. All other PDEs are
first tackled through separation of variables.
\subsection{The {\it heuristic} algorithm for separating the variables}
\label{heuristic}
The main idea of the method of separation of variables consists of
building a system of {\it uncoupled} ODEs equivalent to the original PDE.
To determine the appropriate submethod to separate the variables, we
implemented a {\it heuristic} algorithm, i.e. one that elaborates an
ansatz whose success cannot be determined {\it a priori}. Concerning the
algorithm itself, it seemed reasonable to us to build the ansatz with
selected terms of the general sum of possible products, given, for
example, for an indeterminate function $f(x,y,z)$, by
$$
f_1(x)+f_2(y)+f_3(z)+f_1(x)f_2(y)+f_1(x)f_3(z)
+f_2(y)f_3(z)+f_1(x)f_2(y)f_3(z) \label{general}
$$
For instance, separation {\it by sum} and {\it by product} would be
obtained by selecting only the first three terms or just the last term,
respectively, from the above; the criteria for selecting the terms depend
on the structure of the given PDE.
When the ansatz leads to a partial separation of variables, {\bf pdsolve}
was programmed to reenter itself with the part containing the
non-separated variables as argument, applying the whole strategy once
again to that part.  This usually results in a mixed combination of
separation of variables and {\it deterministic} methods to solve a single
given PDE.
\subsection{The {\tt HINT} option of {\bf pdsolve}}
\label{hint}

As mentioned before, there is no general, ever-efficient method to solve
PDEs. To minimize this problem, our idea was to permit an active
participation of the user in the solving process. This was obtained by
designing {\bf pdsolve} so as to always follow a specific instruction
concerning how to tackle the received PDE. We called this instruction
\verb-HINT-, and it can be given by the user; otherwise, it will be
automatically generated by one of the subroutines of {\bf pdsolve}. In
both cases, the \verb-HINT- may be an indication of either a solving
method already known to {\bf pdsolve}, or a mathematical expression to be
taken as starting point in searching for the solution. As for the user's
specification of a \verb-HINT-, a noteworthy possibility is that of
proposing a {\it functional} \verb-HINT-; that is, to propose the
indeterminate function as an expression involving functions possibly
depending on many variables each, mapping the original PDE into another
PDE. Functional \verb-HINT-s significantly increase {\bf pdsolve}'s
possibilities of finding a solution using users' advice.
\section{The {\it PDEtools} package} A brief overview of the most relevant
commands of the package together with a few simple examples, which can be
checked by hand, is as follows\footnote{The {\it PDEtools} package can be
found, at any of the Internet addresses of the Maple Share Library, while
tutorial sessions and the version under development can be obtained in the
anonymous FTP site of the Symbolic Computing Group at UERJ: 152.92.4.69.} :

\smallskip \noindent $\bullet$ {\bf pdsolve} looks for the general solution
or the complete separation of the variables of a given PDE.

\noindent Example:
\begin{displaymath}
\begin{array}{l}
\mbox{\it PDE: } \left( \! \,{\frac {{ \partial}}{{ \partial}{y}}}\,{\rm
f}(\,{x}
, {y}, {z}\,)\, \!  \right)
+
{\displaystyle
\frac
{1}{{x}{z}}}{{\frac {{\partial}}{{ \partial}{x}}}\,{\rm f}(\,{x}, {y}, {z}\,)}
={\frac {{ \partial}}{{ \partial}{z}}}\,{\rm f}(\,{x}, {y}, {z}
\,) \\*[.15 in]
\mbox{\it pdsolve's answer: }{\rm f}(\,{x}, {y}, {z}\,)={\rm \_f1}(\,{x}^{2}
+ 2\,{i}\,{ \pi} + 2\,{\rm ln}(\,{z}\,), {z} + {y}\,)
\end{array}
\end{displaymath}

\noindent That is, a general solution was returned in terms of an
arbitrary function, \_f1, with two differential invariants as arguments.

\smallskip \noindent $\bullet$ {\bf dchange} performs changes of variables
in PDEs and other algebraic objects (integro-differential equations,
limits, multiple integrals, etc...). New feature of this version is that
{\bf dchange} can also be used to analyze the underlying invariance groups
of a PDE, since it works with changes of both the independent and the
dependent variables, automatically extending the transformation equations
to any required differential order.

\noindent Example: The most general 1st. order ODE admitting the rotation
group.
\begin{displaymath}
\begin{array}{l}
\mbox{\it ODE:   } {\displaystyle \frac { - {y} + {x}\, \left(
\! \,{\frac {{ \partial}}{{ \partial}{x}}}\,{y}\, \!  \right) }{{
x} + {y}\, \left( \! \,{\frac {{ \partial}}{{ \partial}{x}}}\,{y}
\, \!  \right) }}={\rm H}\! \left( \! \,\sqrt {{x}^{2} + {y}^{2}}\,
 \!  \right)
\\*[.2 in]
\mbox{\it transformation: } \{\,{y}={\rm cos}(\,{ \varepsilon}\,)\,y^{*} -
x^{*}\,{\rm sin}(\,{ \varepsilon}\,),\
{x}=y^{*}\,{\rm sin}(
\,{ \varepsilon}\,) + x^{*}\,{\rm cos}(\,{ \varepsilon}\,)\,\}
\\*[.13 in]
\mbox{\it dchange's answer: }
{\displaystyle \frac { - y^{*} + x^{*}\, \left( \! \,
{\frac {{ \partial}}{{ \partial}x^{*}}}\,y^{*}\, \!
 \right) }{x^{*} + y^{*}\, \left( \! \,{\frac {{ \partial}
}{{ \partial}x^{*}}}\,y^{*}\, \!  \right) }}={\rm H}\!
 \left( \! \,\sqrt {y^{*\,2} + x^{*\,2}}\, \!  \right)
\\*[.17 in]
\end{array}
\end{displaymath}
Introducing canonical coordinates $(r,\theta)$, the ODE is reduced to a
quadrature:
\begin{displaymath}
\begin{array}{l}
\mbox{\it transformation: }
\{\,{x}={r}\,{\rm cos}(\,{ \theta}\,),\
{y}={r}\,{\rm sin}(\,{ \theta}\,)\,\}
\\*[.13 in]
\mbox{\it dchange's answer: }
{r} \left( \! \,{\frac {{ \partial}}{{ \partial}{r}}}\,{ \theta
}\, \!  \right) ={\rm H}\! \left( \! \,{r}\, \!
 \right)
\end{array}
\end{displaymath}
$\bullet$ {\bf sdsolve} looks for a complete (or
partial) solution of a {\it coupled} system of ODEs (linear or not), by
mixing linear algebra techniques with a {\it sequential} approach:

\noindent Example:
\begin{displaymath}
\mbox{\it ODEs system: }
\left\{ \!
{\frac {{ \partial}^{2}}{{ \partial}{x}^{2}}}\,{\rm h}(\,{x}\,)
    ={\displaystyle \frac {{\rm g}(\,{x}\,)}{{\rm f}(\,{x}\,)}},\ \ \
{\frac {{ \partial}}{{ \partial}{x}}}\,{\rm g}(\,{x}\,)
    = - {\rm f}(\,{x}\,),\ \ \
{\frac {{ \partial}}{{\partial}{x}}}\,{\rm f}(\,{x}\,)
    ={\rm e}^{{\rm g}(\,{x}\,)}\, \!
 \right\}
\end{displaymath}
\begin{displaymath}
\begin{array}{l}
\mbox{\it sdsolve's answer: }
\! \!  \left[ {\vrule height1.17em width0em depth1.17em}
 \right. \! \!  \left\{ \! \,{\rm f}(\,{x}\,)={\rm tanh} \left(
\! \,{\displaystyle \frac {1}{2}}\,\sqrt {{\it \_C3}}\,(\,{x} +
{\it \_C4}\,)\,\sqrt {2}\, \!  \right) \,\sqrt {2}\,\sqrt {{\it
\_C3}}\, \!  \right\} ,   \\
\left\{ \!\, {\rm g}(\,{x}\,)={\rm ln}
\left( \! {\frac {{ \partial}}{{ \partial}{x}}}\,{\rm f}(\,{x}
\,)\, \!  \right)  \! \, \right\} ,
\left\{ \! \,{\rm h}(\,{x}\,)= - {\displaystyle \int}
{\displaystyle \frac {{\rm g}(\,{x}\,){x}}{{\rm f}(\,{x}\,)}}\,
{d}{x} + {\displaystyle \int} {\displaystyle \frac {{\rm g}(\,{x}
\,)}{{\rm f}(\,{x}\,)}}\,{d}{x}\,{x} + {\it \_C1} + {\it \_C2}
{x}\, \!  \right\}
\! \! \left. {\vrule height1.17em width0em depth1.17em} \right]
\end{array}
\end{displaymath}
$\bullet$ {\bf mapde} maps PDEs into other PDEs, more
convenient in some cases. The mappings implemented and tested up to now
are:
\begin{itemize}
\vspace{-2mm}\item PDEs that explicitly depend on the indeterminate
function into PDEs which do not (i.e. only depend on it through
derivatives).
\item linear $2^{nd}$ order PDEs, with two differentiation
variables, into PDEs in canonical form. Though the success depends on the
proposed problem, in principle this program works both with PDEs with
constant or variable coefficients\footnote{In the case of variable
coefficients, this kind of mapping involves the solving of auxiliary
PDEs.}.
\end{itemize} \vspace{-2mm}

\noindent Example: mapping of a $2^{nd}$ order PDE into canonical form.
\begin{displaymath}
\begin{array}{l}
{\mbox{\it PDE: }} {x}^{2}\, \left( \! \,{\frac {{ \partial}^{2}}{{
\partial}{x}^{2}}}\,{\rm f}(\,{x}, {y}\,)\, \!  \right)  + 2\,{y}
\,{x}\, \left( \! \,{\frac {{ \partial}^{2}}{{ \partial}{y}\,{
\partial}{x}}}\,{\rm f}(\,{x}, {y}\,)\, \!  \right)  + {y}^{2}\,
 \left( \! \,{\frac {{ \partial}^{2}}{{ \partial}{y}^{2}}}\,{\rm
f}(\,{x}, {y}\,)\, \!  \right)=0
\\*[.13in]
\mbox{\it mapde's answer: }
 \left( \! \,{\frac {{ \partial}^{2}}{{ \partial}{\_ \xi 1}^{2}}}
\,{\rm f}(\,{\_ \xi 1}, {\_ \xi 2}\,)\, \!  \right) \,{\_ \xi 1}
^{2}=0,\, \, \, \,
{\rm where}\, \left\{ \! \,{\_ \xi 2}= - \,
{\displaystyle \frac {{y}}{{x}}}, {\_ \xi 1}={x}\, \!  \right\}
\end{array}
\end{displaymath}

\smallskip \noindent $\bullet$ {\bf strip} evaluates the {\it characteristic
strip} associated to a first-order PDE.

\smallskip \noindent $\bullet$ {\bf pdtest} tests a solution found by {\bf
pdsolve} for a given PDE by making a careful simplification of the PDE with
respect to this solution.
\section*{Conclusions}
Taking into account the algorithmic character of many of the existing
exact solving methods for PDEs, it is clear that computational approaches
can play an important role in the progress in this area. In this context,
the goal of the material here presented is to contribute both to the
discussion of what would be a computational strategy for the problem and
to the implementation of that strategy using the mathematical methods and
the algebraic computing resources available nowadays. In particular,
considering there is still no general method able to solve {\it all}
possible PDEs, great emphasis was put in the {\it interactive} character
of the package  (the \verb-HINT- option and the {\bf dchange} command). We
believe this to be a basic feature any PDE-solving package should have.

Finally, routines for the analytical solving of systems of PDEs are now
under development. We expect to report them in the near future.
\section*{Acknowledgments}
This work was supported by the State University of Rio de Janeiro (UERJ)
and the National Research Council (CNPq) of Brazil.
\end{document}